\documentclass[12pt,groupcitations]{article}
\pdfoutput=1
\usepackage[T1]{fontenc}
\usepackage[ansinew]{inputenc}
\usepackage[italian,english]{babel}
\usepackage{amsfonts}
\usepackage{amsmath}
\usepackage{array}
\usepackage{amsthm}
\usepackage{amssymb}
\pdfoutput=1
\usepackage{graphicx}
\usepackage{subfigure}
\usepackage{braket}
\usepackage{eucal}
\usepackage{verbatim}
\usepackage{caption}
\usepackage{multicol}
\usepackage{tikz}

\usepackage{float}

\usepackage{multirow}

\usetikzlibrary{arrows,shapes}
\usetikzlibrary{matrix,arrows}
\newcommand{\be}{\begin{eqnarray}}
\newcommand{\ee}{\end{eqnarray}}
\begin{document}
\title{\bf Quasinormal Modes of Modified Gravity (MOG) Black Holes}

%
% \bigskip
%
 %\centerline
\author{{Luciano Manfredi$^{a,}$\footnote{\textit{E-mail:} \texttt{lmanfred@lion.lmu.edu}},  Jonas Mureika $^{a,}$\footnote{\textit{E-mail:} \texttt{jmureika@lmu.edu}}, John Moffat$^{b,}$\footnote{\textit{E-mail:} \texttt{jmoffat@perimeterinstitute.ca}}}
\bigskip\\
\bigskip\\
\textit{$^a$Department of Physics, Loyola Marymount University, Los Angeles, CA  90045-2659} \\
\textit{$^b$Perimeter Institute for Theoretical Physics, Waterloo, Ontario N2L 2Y5, Canada}}

\maketitle

\begin{abstract}
The Quasinormal modes (QNMs) for gravitational and electromagnetic perturbations are calculated in a Scalar-Tensor-Vector (Modified Gravity) spacetime, which was initially proposed to obtain correct dynamics of galaxies and galaxy clusters without the need for dark matter. It is found that for the increasing model parameter $\alpha$, both the real and imaginary parts of the QNMs decrease compared to those for a standard Schwarzschild black hole.  On the other hand, when taking into account the $1/(1+\alpha)$ mass re-scaling factor present in MOG, Im($\omega$) matches almost identically that of GR, while Re($\omega$) is higher. These results can be identified in the ringdown phase of massive compact object mergers, and are thus timely in light of the recent gravitational wave detections by LIGO. 
\end{abstract}
\pagebreak
\section{Introduction}

We are at the dawn of the era of gravitational wave astronomy.  The four binary black hole merger signals observed by LIGO since September 2015 -- 
GW150914~\cite{Abbott:2016blz}, GW-151226~\cite{Abbott:2016nmj}, GW-170104~\cite{Abbott:2017vtc} and GW170814 \cite{Abbott:2017oio} -- not only serve as concrete evidence of the existence of black holes, but allow us to explore classical (and potentially quantum) gravitational physics in a new fashion.  A binary neutron star merger has also been observed (GW170817, \cite{TheLIGOScientific:2017qsa}), giving more data for non-black compact objects.  Gravitational waves also present a novel test of the underlying theory of gravitation, since their characteristics depend uniquely on the background spacetime structure.  Such data can thus be a testbed for general relativity (GR)~\cite{TheLIGOScientific:2016src}, as well as all alternate theories of gravitation ~\cite{Konoplya:2011qq,Sibandze:2017jbi,Lasky:2010bd,Bhattacharyya:2017tyc}.

A binary merger signal is marked by three phases: the inspiral, merger, and ringdown.  The characteristics of each phase yield important information about the nature of the merging objects (size, mass, spin, {\it etc...}).  The inspiral phase is defined by the chirp frequency, and the associated waveform can be calculated for a given spacetime structure.  The ringdown phase -- in which the final black hole horizon ``settles'' down through damped oscillations -- is in turn characterized by quasinormal mode vibrations (QNMs)~\cite{Kokkotas:1999bd}.  The QNM distribution can be obtained from the ringdown phase of the merger, and can provide important insight into the process.  

It is likely that QNMs can be deciphered from such experimental data either in LIGO~\cite{Goggin:2008dz,Abbott:2009km,Aasi:2014bqj}, or eventually in LISA \cite{Berti:2005ys}.  Identification of model-dependent characteristics is thus of timely importance, including generic spacetime dependence, but also aspects such as the onset time for QNMs in mergers \cite{Sakai:2017ckm}, and higher-order modes~\cite{London:2014cma}.
It has recently been pointed out that such data can also be useful in differentiating the type of compact objects that have merged ~\cite{Cardoso:2016rao}.
 
In this paper, we calculate quasinormal mode frequencies for static black holes in a Scalar-Tensor-Vector Gravity (STVG) gravity theory~\cite{Moffat:2006,Moffat:2016,Moffat:2014aja,Moffat:2014,Moffat:2017} using the Asymptotic Iteration Method (AIM)~\cite{Cho:2011sf,qnmaim}.  We begin by reviewing the essentials of MOG, QNMs, and the AIM, after which we present the $\ell=2,3,4$ frequencies of gravitational and $\ell = 1,2,3$ electromagnetic perturbations.  The QNMs reduce to the standard values in the limit of general relativity.  Lastly, we discuss the distinctions between MOG QNMs and GR QNMs, and consider the experimental detection of QNMs in either LIGO or LISA data, and the possibility of distinguihing GR from MOG QNMs.

\section{Scalar-Tensor-Vector Modified Gravity (MOG)}

In the following, we will investigate the ringdown phase of merging black holes producing gravitational waves in the generalized theory of gravitational STVG (MOG)~\cite{Moffat:2006}. The theory has been studied as an alternative to GR without detectable dark matter in the present universe and fits to galaxy rotation curves and galaxy clusters have been obtained~\cite{MoffatRahvar1,MoffatRahvar2,MoffatToth}. The theory has also been applied to cosmology explaining early universe structure growth
and the CMB data~\cite{Moffat3,Moffat4,Shojai}.  

The field equations in STVG are given by (c=1):
\begin{equation} 
\label{mog1}
G_{\mu\nu}-\Lambda g_{\mu\nu}+Q_{\mu\nu}=-8\pi GT_{\mu\nu},
\end{equation}
\begin{equation}
\label{mog2} 
\frac{1}{\sqrt{-g}}\partial_{\nu}(\sqrt{-g}B^{\mu\nu})+ \mu^2\phi^\mu=-J^\mu,
\end{equation}
\begin{equation}
\label{mog3}
\partial_\sigma B_{\mu\nu}+\partial_\mu B_{\nu\sigma}+\partial_\nu B_{\sigma\mu}=0,
\end{equation}
\begin{equation}
\label{mog4}
\Box G=K(x),
\end{equation}
\begin{equation}
\label{mog5}
\Box\mu=L(x).
\end{equation}
We have
\begin{equation}
Q_{\mu\nu}=\frac{2}{G^2}(\partial^\alpha G\partial_\alpha Gg_{\mu\nu}-\partial_\mu G\partial_\nu G)-\frac{1}{G}(\Box Gg_{\mu\nu}-\nabla_\mu\partial_\nu G).
\end{equation}
Moreover, 
\begin{equation}
K(x)=\frac{3}{G}\biggl(\frac{1}{2}\partial^\alpha G\partial_\alpha G-V(G)\biggr)-\frac{3}{G}\partial^\alpha G\partial_\alpha G
+\frac{G}{\mu^2}\biggl(\frac{1}{2}\partial^\alpha\mu\partial_\alpha\mu-V(\mu)\biggr)+\frac{3G^2}{16\pi}\Box\biggl(\frac{1}{G}\biggr),
\end{equation} 
and
\begin{equation}
L(x)=-\biggl(\frac{1}{G}\partial^\alpha G\partial_\alpha\mu+\frac{2}{\mu}\partial^\alpha\mu\partial_\alpha\mu-\mu^2G\frac{\partial V(\phi_\mu))}{\partial\mu}\biggr),
\end{equation}
where $V(\phi_\mu)=(1/2)\mu^2\phi^\mu\phi_\mu$. Moreover, $G_{\mu\nu}$ is the Einstein tensor $G_{\mu\nu}=R_{\mu\nu}-\frac{1}{2}g_{\mu\nu}R$, $\Lambda$ is the cosmological constant, $\nabla_\mu$ is the covariant derivative with respect to $\Gamma^\lambda_{\mu\nu}$, $\Box=\nabla^\mu\nabla_\mu$, $T=g^{\mu\nu}T_{\mu\nu}$, $G$ and $\mu$ are scalar fields and $V(G)$ and V$(\mu)$ are potentials\footnote[1]{Eq. (8) in ~\cite{Moffat:2016} should be replaced by Eq. (\ref{mog3}).}. The Ricci curvature tensor is defined by
\begin{equation}
R_{\mu\nu}=\partial_\nu{\Gamma_{\mu\sigma}}^\sigma-\partial_\sigma{\Gamma_{\mu\nu}}^\sigma+\Gamma_{\mu\sigma}^\alpha\Gamma_{\alpha\nu}^\sigma
-\Gamma_{\mu\nu}^\alpha\Gamma_{\alpha\sigma}^\sigma.
\end{equation}

The total energy-momentum tensor is 
\begin{eqnarray} 
T_{\mu\nu}=T^M_{\mu\nu}+T^\phi_{\mu\nu}+T^G_{\mu\nu}+T^\mu_{\mu\nu},
\end{eqnarray} 
where $T^M_{\mu\nu}$ is the energy-momentum tensor for ordinary matter, and
\begin{equation}
\label{mog7}
T^{(\phi)}_{\mu\nu}=-\frac{1}{4\pi}\biggl[B_{\mu}^{~\alpha}B_{\nu\alpha}-g_{\mu\nu}\left(\frac{1}{4}B^{\rho\alpha}B_{\rho\alpha}+V(\phi_\mu)\right)
+2\frac{\partial V(\phi_\mu)}{\partial g^{\mu\nu}}\biggr],
\end{equation}
\begin{equation}
T^{(G)}_{\mu\nu}=-\frac{1}{4\pi G^3}\biggl(\partial_\mu G\partial_\nu G-\frac{1}{2}g_{\mu\nu}\partial_\alpha G\partial^\alpha G\biggr),
\end{equation}
\begin{equation}
T^{(\mu)}_{\mu\nu}=-\frac{1}{4\pi G\mu^2}\biggl(\partial_\mu\mu\partial_\nu\mu-\frac{1}{2}g_{\mu\nu}\partial_\alpha\mu\partial^\alpha\mu\biggr).
\end{equation}

The covariant current density $J^\mu$ for matter is defined by
\begin{equation}
\label{currentdensity}
J^\mu=\kappa T^{M\mu\nu}u_\nu,
\end{equation}
where $\kappa=\sqrt{\alpha G_N}$, $\alpha=(G-G_N)/G_N$ is a dimensionless scalar field, $G_N$ is Newton's constant, $u^\mu=dx^\mu/ds$ and $s$ is the proper time along a particle trajectory.  The perfect fluid energy-momentum tensor for matter is given by
\begin{equation}
\label{energymomentum}
T^{M\mu\nu}=(\rho_M+p_M)u^\mu u^\nu-p_Mg^{\mu\nu},
\end{equation}
where $\rho_M$ and $p_M$ are the density and pressure of matter, respectively, and for the fluid $u^\mu$ is the comoving four-velocity. We get from (\ref{currentdensity}) and (\ref{energymomentum}) by using $u^\nu u_\nu=1$:
\begin{equation}
J^\mu=\kappa\rho_M u^\mu.
\end{equation}
The gravitational source charge is given by
\begin{equation}
Q_g=\int d^3x J^0(x).
\end{equation}

\section{MOG Black Holes}

An exact generalized Schwarzschild-MOG solution of the STVG fields equations is obtained by requiring that $G=G_N(1+\alpha)\sim$ constant and $Q_g=\sqrt{\alpha G_N}M\sim$ constant, and ignoring the small $\phi_\mu$ field particle mass $m_\phi\sim 10^{-28}$ eV in the present universe. This mass is obtained from fitting STVG to the galaxy and cluster data~\cite{MoffatRahvar1},~\cite{MoffatRahvar2}. The field equations are given by
\begin{equation}
\label{phiFieldEq}
R_{\mu\nu}=-8\pi GT^\phi_{\mu\nu},
\end{equation}
\begin{equation}
\label{Bequation}
\frac{1}{\sqrt{-g}}\partial_\nu(\sqrt{-g}B^{\mu\nu})=0,
\end{equation}
\begin{equation}
\label{Bcurleq}
\partial_\sigma B_{\mu\nu}+\partial_\mu B_{\nu\sigma}+\partial_\nu B_{\sigma\mu}=0.
\end{equation}
The energy-momentum tensor ${{T^{(\phi)}}_\mu}^\nu$ is 
\begin{equation}
\label{Tphi}
{{T^{(\phi)}}_\mu}^\nu=-\frac{1}{4\pi}({B_{\mu\alpha}}B^{\nu\alpha}-\frac{1}{4}{\delta_\mu}^\nu B^{\alpha\beta}B_{\alpha\beta}).
\end{equation}

The metric is given by
\be
\label{MOGSchwarz}
ds^2=&\left(1-\frac{2G_N(1+\alpha)M}{r}+\frac{\alpha(1+\alpha)G_N^2M^2}{r^2}\right)dt^2\\
&-\left(1-\frac{2G_N(1+\alpha)M}{r}+\frac{\alpha(1+\alpha)G_N^2M^2}{r^2}\right) ^{-1}dr^2-r^2d\Omega^2.
\ee
This has the form of the static, spherically symmetric point particle Reissner-Nordstrom solution for an electrically charged black hole, but now the charge $Q_g > 0$ is of gravitational origin. As in the case of astrophysical bodies, black holes are not expected to possess electric charge.

The MOG black hole possesses two horizons given by
\be
r_\pm=G_NM\left[1+\alpha \pm \left(1+\alpha\right) ^{1/2}\right].
\ee	
Note that we obtain the Schwarzschild black hole when $\alpha=0$ giving $r_+=r_s=2G_NM$ as expected. 

A generalized Kerr-MOG black hole solution has also been derived~\cite{Moffat:2014aja}. The metric is 
\begin{equation}
\label{KerrMOG}
ds^2=\frac{\Delta}{\rho^2}(dt-a\sin^2\theta d\phi)^2-\frac{\sin^2\theta}{\rho^2}[(r^2+a^2)d\phi-adt]^2-\frac{\rho^2}{\Delta}dr^2-\rho^2d\theta^2,
\end{equation}
where
\begin{equation}
\Delta=r^2-2GMr+a^2+\alpha(1+\alpha) G_N^2M^2,\quad \rho^2=r^2+a^2\cos^2\theta.
\end{equation}
The spacetime geometry is axially symmetric around the $z$ axis. Horizons are determined by the roots of $\Delta=0$:
\begin{equation}
r_\pm=G_N(1+\alpha)M\biggl[1\pm\sqrt{1-\frac{a^2}{G_N^2(1+\alpha)^2M^2}-\frac{\alpha}{1+\alpha}}\biggr].
\end{equation}
An ergosphere horizon is determined by $g_{00}=0$:
\begin{equation}
r_E=G_N(1+\alpha)M\biggl[1+\sqrt{1-\frac{a^2\cos^2\theta}{G_N^2(1+\alpha)^2M^2}-\frac{\alpha}{1+\alpha}}\biggr].
\end{equation}
The solution is fully determined by the Arnowitt-Deser-Misner (ADM) mass $M$ and spin parameter $a$ ($a=S/GM$ where $S$ denotes the spin-angular momentum) measured by an asymptotically distant observer. When $a=0$ the solution reduces to the Schwarzschild MOG black hole metric (\ref{MOGSchwarz}).

\section{Calculating Quasinormal Modes}
Investigations concerning the interaction of black holes with surrounding fields give us the possibility to learn about black hole physics. Some of this information can be obtained from QNMs, which are characteristic to the background black hole's space-time. QNMs present complex frequencies, whose real part gives the actual frequency while the imaginary part dictates the damping.
	
The QNM gravitational perturbations are considered to be the most important type of perturbations to be analyzed, since they can directly identify black holes and their gravitational radiation. In other words, gravitational QNMs serve as a unique fingerprint when searching for the existence of black holes. With the newly launched era of gravitational wave astronomy, triggered by the detection of a transient gravitational-wave signal determined to be the coalescence of two black holes back on September 14th, 2015, the understanding of QNMs grew in popularity, and even more for alternative theories of gravity like the one examined in this paper.

Formally, QNMs are solutions to perturbed gravitational field equations subject to the two boundary conditions
\be
\psi(x)\longrightarrow \left\lbrace \begin{array}{cc}
e^{-i \omega x} & x \rightarrow -\infty\\
e^{i \omega x} & x \rightarrow  \infty
\end{array} \right.
\ee
where the positive and negative solutions refer to ingoing and outgoing waves, respectively.

Calculation of these quantities often requires a numerically-intensive procedure, and furthermore the boundary conditions listed above are not satisfied in a wide range of cases.  This leads one to identify the QNMs as the discrete values where the above holds true.

With the emergence of accessible large-scale computing over the last two decades, several new methods have emerged as viable avenues to obtaining QNMs.  Among the most popular ones, there is the semi-analytic formalism employed by Ferrari and Mashoon~\textit{et al.}~\cite{mashhoon}, the Continued Fraction Method (CFM) developed by Leaver~\cite{Leaver}, and the WKB approximation \cite{WKB}. In recent years, the asymptotic iteration method (AIM) was shown to be more efficient in certain cases \cite{deaim}.
\subsection{The Asymptotic Iteration Method}
 The AIM we adopt to calculate QNMs is detailed in \cite{Cho:2011sf,qnmaim}, and will be the machinery of focus in this paper.  We reproduce the essential components of the method here.

We begin by defining a second-order differential equation of the form \be
\chi''=\lambda_{0}(x)\chi'+s_{0}(x)\chi~~.
\ee
The functions $\lambda_{0}(x)$ and $s_{0}(x)$ are well defined and sufficiently smooth, which allows us to further differentiate the above to obtain
\be
\chi'''=\lambda_{1}(x)\chi'+s_{1}(x)\chi~~.
\ee
Here, the new coefficients are $\lambda_{1}(x)=\lambda_{0}'+s_{0}+\lambda_{0}^2$ and $s_{1}(x)=s_{0}'+s_{0}\lambda_{0}$.
\vspace{0.03in}

Proceeding from an iterative approach in the order of differentiation, one eventually arrives at the general expression
\be 
\chi^{(n+2)}=\lambda_{n}(x)\chi'+s_{n}(x)\chi,
\ee
where the coefficients satisfy the relations
\be\label{recrel}
\lambda_{n}(x)=\lambda_{n-1}'+s_{n-1}+\lambda_{0}\lambda_{n-1}, \ \  \   \ s_{n}(x)=s_{n-1}'+s_{0}\lambda_{n-1}.
\ee

For $n$ large enough, the AIM feature is manifested by requiring 
\be
\frac{s_n(x)}{\lambda_n(x)}=\frac{s_{n-1}(x)}{\lambda_{n-1}(x)}\equiv \beta (x),
\ee
The QNMs thus arise from a "quantization condition" marking an end to the algorithm \cite{termination},
\be\label{quant}
\delta_{n}=s_n\lambda_{n-1}-s_{n-1}\lambda_n=0.
\ee

Nevertheless, these recursion relations (\ref{recrel}) are not numerically optimal to evaluate ~\cite{recrelimprovement}, since differentiation of the $\lambda$ and $s$ terms in previous steps needs to be taken repetitively. The parameters  $\lambda_n$ and $s_n$ are Taylor-expanded about the point $\xi$,
\be
\lambda_n(\xi)=\sum_{i=0}^{\infty}c_n^i(x-\xi)^i\\
s_n(\xi)=\sum_{i=0}^{\infty}d_n^i(x-\xi)^i,
\ee
with the values $c_n^i$ and $d^i_n$ representing the $i^{th}$ Taylor constants of $\lambda_n(\xi)$ and $s_n(\xi)$. 
A new recursive expression is obtained after substituting the above values into (\ref{recrel}),
\be
&c_n^i=(i+1)c_{n-1}^{i+1}+d^i_{n-1}+\sum_{k=0}^i c_0^k c_{n-1}^{i-k}\\
&d_n^i=(i+1)d_{n-1}^{i+1}+\sum_{k=0}^i d_0^k c_{n-1}^{i-k}.
\ee
Now the quantization condition~(\ref{quant}) is cast in the new form as:
\be
d_n^0c_{n-1}^0-d_{n-1}^0c_n^0=0,
\ee
which saves computational time \cite{qnmaim}.

%\pagebreak
\section{Quasinormal Modes of Static MOG Black Holes} 
After rescaling $2M = 1$ in natural units $G_N=c=1$ makes QNMs dependent on $Q_g$, $\ell$, and $n$, the new metric function can be written
\be
f(r)=1-\frac{1+\alpha}{r}+\frac{\alpha(1+\alpha)}{4r^2}.
\ee
Outside the event horizon, STVG MOG perturbation equations are separable and produce the even ($+$) and odd ($-$) parity oscillations \cite{Cho:2011sf,qnmaim,parity1,parity2,parity3},
\be\label{qnmeqn}
\left( \frac{d^2}{dx^2}-\rho^2-V_i^{(\pm)}\right)Z_i^{(\pm)}=0,
\ee
with the following quantities defined for $i=j=1,2$, and $i\neq j$,
\be 
&V_i^{(-)}(r)=\frac{\Delta}{r^5}\left(Ar-q_j+\frac{4Q_g^2}{r}\right)\\
&V_i^{(+)}(r)=V_i^{(-)}(r)+2q_j\frac{d}{dx}\left(\frac{\Delta}{r^2\left[(\ell-1)(\ell+2)r+q_j\right]}\right)
\ee
and finally
\be
&\frac{dr}{dx}=\frac{\Delta}{r^2}\label{randx},\\
&\Delta=(r-r_+)(r-r_-)=\frac{1}{4}[4 r^2 - 4 r (1+\alpha) + \alpha(1+\alpha)],\\
&A=\ell(\ell+1),\\
&q_1=\frac{1}{2}\left[3+\sqrt{9+16Q_g^2(\ell-1)(\ell+2)}\right],\\
&q_2=\frac{1}{2}\left[3-\sqrt{9+16Q_g^2(\ell-1)(\ell+2)}\right],\\
&\rho=-i\omega~~~.
\ee
In the above expressions, $\omega$ denotes frequency, $\ell$ the rotational parameter and $r_-$ and $r_+$ correspond to the two horizon radii. %In the $\alpha=0$ limit, $r_+=1$ and $r_-=0$ as expected.

Because the effective potential $V_i^{(-)}$ is easier to deal with than $V_i^{(+)}$, QNMs are calculated for the odd-parity modes. The isospectrality between the QNMs of the odd and even perturbations was numerically determined to hold. Indeed, the two potentials $V_i^{(-)}$ and $V_i^{(+)}$ share the same spectra of QNMs.

The tortoise coordinate $x$ varies between $(-\infty,\infty)$ starting at the horizon off to infinity, and can be written
\be
x=\int\frac{r^2}{\Delta}=r+\frac{r_+^2}{r_+-r_-}\ln(r-r_+)-\frac{r_-^2}{r_+-r_-}\ln(r-r_-),
\ee
Under a change of variable $x \rightarrow r$ using (\ref{randx}), the derivative operators are
\be
\frac{d}{dx}=\frac{\Delta}{r^2}\frac{d}{dr}
\ee
and
\be
\frac{d^2}{dx^2}=\left(\frac{\Delta}{r^2}\right)\left(\frac{(2r-\alpha)(1+\alpha)}{2r^3}\right)\frac{d}{dr}+\left(\frac{\Delta}{r^2}\right)^2\frac{d^2}{dr^2}.
\ee

Re-evaluating (\ref{qnmeqn}) gives
\be\label{qnmeqnr}
\left(\frac{\Delta}{r^2}\right)^2\frac{d^2}{dr^2}Z_i^{(-)}+\left(\frac{\Delta}{r^2}\right)\left(\frac{(2r-\alpha)(1+\alpha)}{2r^3}\right)\frac{d}{dr}Z_i^{(-)}-\left[\rho^2+V_i^{(-)}\right]Z_i^{(-)}=0
\ee
where $Z_i^{(-)}$ satisfies the boundary conditions: 
\be
Z_i^{(-)}\longrightarrow \left\lbrace \begin{array}{cc}
e^{-i \omega x} &, x \rightarrow -\infty\\
e^{i \omega x} &, x \rightarrow  \infty
\end{array} \right.
\ee
The asymptotic behavior is introduced by imposing the previous condition on $Z_i^{(-)}$ \cite{asymbehavior}, giving 
\be
Z_i^{(-)}=e^{-\rho r}r^{-1}(r-r_1)^{1-\rho-\frac{\rho r_+^2}{r_+-r_-}}(r-r_+)^{\frac{\rho r_+^2}{r_+-r_-}}\chi_{Z_i(r)}.
\ee

Taking the $r$ derivative of $Z_i^{(-)}$ once and twice leads to
\be\label{r}
Z_{i,r}^{(-)}\equiv \frac{d}{dr}Z_{i}^{(-)}=e^{-\rho r}r^{-1}(r-r_1)^{1-\rho-\frac{\rho r_+^2}{r_+-r_-}}(r-r_+)^{\frac{\rho r_+^2}{r_+-r_-}}\left(\chi_{Z_{i,r}}+\Gamma_Z \chi_{Z_{i}}\right)
\ee
and
\be\label{rr}
Z_{i,rr}^{(-)}=e^{-\rho r}r^{-1}(r-r_1)^{1-\rho-\frac{\rho r_+^2}{r_+-r_-}}(r-r_+)^{\frac{\rho r_+^2}{r_+-r_-}}\left(\chi_{Z_{i,rr}}+2\Gamma_Z, \chi_{Z_{i,r}}+(\Gamma_Z^2+\Gamma_{Z,r})\chi_{Z_{i}}\right)
\ee
where
\be
\Gamma_Z=-\rho-\frac{1}{r}+\frac{(1-\rho)(r_+-r_-)-\rho r_+^2}{(r_+-r_-)(r-r_-)}+\frac{\rho r_+^2}{(r_+-r_-)(r-r_+)}.
\ee
Evaluating Equation~(\ref{qnmeqnr})  using the expressions (\ref{r}) and (\ref{rr}) leads to 
%Replacing (\ref{r}) and (\ref{rr}) into Eqn. (\ref{qnmeqnr})  gives
\be
&\left( \frac{\Delta}{r^2} \right) ^2 \chi_{Z_{i,rr}}+\left[ 2\Gamma_Z \left( \frac{\Delta}{r^2} \right)^2+\left( \frac{\Delta}{r^2} \right) \left( \frac{(2r-\alpha)(1+\alpha)}{2r^3} \right)\right] \chi_{Z_{i,r}} + \\
&\left\lbrace \left( \frac{\Delta}{r^2}\right)^2 \left(\Gamma_Z^2+\Gamma_{Z,r} \right)+\left( \frac{\Delta}{r^2} \right) \left(\frac{(2r-\alpha)(1+\alpha)}{2r^3} \right)\Gamma_Z-\left[\rho^2+V_i^{(-)}\right] \right\rbrace  \chi_{Z_i}=0.
\ee
Following this step, a change of coordinates is performed such that $r \rightarrow \xi$ with $\xi=1-\frac{r_+}{r}$, which ranges between $[0,1]$ for $r$ from $[r_+,\infty)$.  It follows that the radial derivatives become
\be
\frac{d}{dr}=\frac{(1-\xi)^2}{r_+}\frac{d}{d\xi}
\ee
and
\be
\frac{d^2}{dr^2}=\frac{(1-\xi)^4}{r_+^2}\frac{d^2}{d\xi^2}-2\frac{(1-\xi)^3}{r_+^2}\frac{d}{d\xi}.
\ee
Finally, the AIM formalism can be cast as the combination of the above expressions, which yields
%Plugging these back in and recasting the equation to perform the AIM get
\be
\chi_{Z_{i,\xi\xi}}=\lambda_{Z_{i}}(\xi)\chi_{Z_{i,\xi}}+s_{Z_{i}}(\xi)\chi_{Z_{i}},
\ee
with the functions therein defined as
\be
\lambda_{Z_{i}}(\xi)=\frac{\alpha(\alpha+1) +4 \Delta }{2 \Delta (1-\xi)}-\frac{r_+(\alpha +2 \Gamma_Z
   \Delta +1)}{\Delta  (\xi -1)^2},
\ee
\be
s_{Z_{i}}(\xi)=\frac{r_+^6 \left(\rho ^2+V_i^{(-)}\right)}{\Delta ^2 (\xi -1)^8}-\frac{\Gamma_Z  r_+^2
   (1+ \alpha +\Delta \Gamma_Z)}{\Delta  (\xi -1)^4}-\frac{\alpha  (\alpha +1) \Gamma_Z  r_+}{2 \Delta (\xi -1)^3}-\frac{r_+ \Gamma_{Z,\xi}}{(\xi-1)^2},
\ee
\be
\Gamma_Z=-\rho-\frac{1-\xi}{r_+}+\frac{\left[(1-\rho)(r_+-r_-)-\rho r_+^2 \right](1-\xi)}{(r_+-r_-)\left[r_+-r_-(1-\xi)\right]}+\frac{\rho r_+(1-\xi)}{(r_+-r_-)\xi},
\ee
\be
V_i^{(-)}=\Delta\frac{(1-\xi)^5}{r^5_+}\left[\frac{A r_+}{1-\xi}-q_j+\frac{4Q_g^2(1-\xi)}{r_+}\right],
\ee
\be
\Delta=\frac{r_+ \xi \left[r_+-r_-(1-\xi)\right]}{(1-\xi)^2}=\frac{(1+\alpha) \xi \left(2+2\sqrt{1+\alpha}+\alpha \xi\right)}{4(\xi-1)^2},
\ee
\be
r_\pm=\frac{1}{2}\left[1+\alpha \pm \left(1+\alpha\right) ^{1/2}\right],
\ee
\be
Q_g=\frac{\sqrt{\alpha}}{2}.
\ee

%\pagebreak

The QNMs obtained are illustrated in Table \ref{Table1} and Table \ref{Table2}, showing the real and imaginary parts of $\omega$, with mass re-scaled back to $M=1$ for easier comparison with literature.  

At the $\alpha=0$ limit, one recovers purely gravitational QNMs for $V_{i=2}^{(-)}$ in Table \ref{Table2} and purely electromagnetic QNMs for $V_{i=1}^{(-)}$ in Table \ref{Table1}. It is worth mentioning that just like in GR, there is no gravitational radiation carried in dipolar perturbations. Because the gravitational charge of the vector field is $Q_g=\sqrt{\alpha G_N }M > 0$ and $M>0$, $Q_g$ is never negative making it impossible for MOG to have dipole radiation \cite{Moffat:2016}. Lastly, we performed a complete scan of the space for all modes with $\ell=1,2,3$ and $\ell=2,3,4$ in both sectors $Z_1$ and $Z_2$ respectively, and no unstable QNMs with Im$(\omega)>0$ were found.

Some comments on the higher $(n=3)$ overtones are perhaps necessary. In general, to ensure accuracy of the results the AIM code was compiled successively for increasing number of iterations until no variation in the value of QNMs was observed. Usually, runs at 400 and 500 iterations yielded the same frequencies in every case, except for $n \gg \ell$ as in Table \ref{Table1} for $\ell=1$ and $n=3$, where the frequencies stabilized at 600 and 700 iterations.

\begin{table}[h!]
\begin{tabular}{ |c|c|c|c|c|c| } 
 \hline
 \hline
 $\ell$ & $n$ & $\alpha = 0$ & $\alpha = 1$ & $\alpha = 4$ & $\alpha = 9$ \\ 
 \hline
 \multirow{4}{1em}{1} & 0 & 0.2483 - 0.09249i & 0.1448 - 0.04805i & 0.06343 - 0.01881i & 0.03268 - 0.009084i \\
   & 1 & 0.2145 - 0.2937i & 0.1308 - 0.1506i & 0.05882 - 0.05828i & 0.03038 - 0.02796i \\
   & 2 & 0.1748 - 0.5252i & 0.1135 - 0.2654i & 0.05258 - 0.1014i & 0.02675 - 0.04833i \\
   & 3 & 0.1462 - 0.7719i & 0.1090 - 0.3866i & 0.05036 - 0.1494i & 0.02434 - 0.07008i \\
 \hline
 \multirow{4}{1em}{2} & 0 & 0.4576 - 0.09500i & 0.2651 - 0.04917i & 0.1164 - 0.01930i & 0.06000 - 0.009351i \\
   & 1 & 0.4365 - 0.2907i & 0.2565 - 0.1498i & 0.1136 - 0.05854i & 0.05861 - 0.02830i \\
   & 2 & 0.4012 - 0.5016i & 0.2420 - 0.2563i & 0.1087 - 0.09948i & 0.05607 - 0.04791i \\
   & 3 & 0.3626 - 0.7302i & 0.2257 - 0.3699i & 0.1028 - 0.1425i & 0.05272 - 0.06840i \\
 \hline
 \multirow{4}{1em}{3} & 0 & 0.6569 - 0.09562i & 0.3771 - 0.04933i & 0.1648 - 0.01936i & 0.08493 - 0.009395i \\
   & 1 & 0.6417 - 0.2897i & 0.3709 - 0.1492i & 0.1627 - 0.05842i & 0.08391 - 0.02831i \\
   & 2 & 0.6138 - 0.4921i & 0.3594 - 0.2522i & 0.1589 - 0.09841i & 0.08198 - 0.04760i \\
   & 3 & 0.5779 - 0.7063i & 0.3446 - 0.3600i & 0.1537 - 0.1398i & 0.07927 - 0.06743i \\
 \hline
 \hline
\end{tabular}
\caption{QNMs accurate to 4 decimal places for $M=1$ scaled MOG electromagnetic perturbations $V_{i=1}^{(-)}$ for $\ell=1$, $\ell=2$ and $\ell=3$ modes.}
\label{Table1}
\end{table}
\begin{table}[h!]
\begin{tabular}{ |c|c|c|c|c|c| } 
 \hline
 \hline
 $\ell$ & $n$ & $\alpha = 0$ & $\alpha = 1$ & $\alpha = 4$ & $\alpha = 9$ \\ 
 \hline
 \multirow{4}{1em}{2} & 0 & 0.3737 - 0.0890i & 0.2220 - 0.04650i & 0.1021 - 0.01867i & 0.05431 - 0.009171i \\
   & 1 & 0.3467 - 0.2739i & 0.2115 - 0.1423i & 0.09872 - 0.05678i & 0.05270 - 0.02781i \\
   & 2 & 0.3011 - 0.4783i & 0.1937 - 0.2457i & 0.09283 - 0.09696i & 0.04974 - 0.04725i \\
   & 3 & 0.2515 - 0.7051i & 0.1742 - 0.3579i & 0.08582 - 0.1397i & 0.04584 - 0.06776i \\
 \hline
 \multirow{4}{1em}{3} & 0 & 0.5994 - 0.0927i & 0.3353 - 0.04758i & 0.1496 - 0.0189i & 0.07875 - 0.009267i \\
   & 1 & 0.5826 - 0.2813i & 0.3281 - 0.1441i & 0.1472 - 0.0571i & 0.07761 - 0.02795i \\
   & 2 & 0.5517 - 0.4791i & 0.3149 - 0.2444i & 0.1428 - 0.0964i & 0.07543 - 0.04706i \\
   & 3 & 0.5120 - 0.6903i & 0.2979 - 0.3503i & 0.1368 - 0.1373i & 0.07238 - 0.06680i \\
 \hline
 \multirow{4}{1em}{4} & 0 & 0.8092 - 0.0942i & 0.4452 - 0.04804i & 0.1965 - 0.01903i & 0.1030 - 0.009311i \\
   & 1 & 0.7966 - 0.2843i & 0.4398 - 0.1449i & 0.1947 - 0.05731i & 0.1021 - 0.02802i \\
   & 2 & 0.7727 - 0.4799i & 0.4294 - 0.2441i & 0.1912 - 0.09625i & 0.1004 - 0.04699i \\
   & 3 & 0.7398 - 0.6839i & 0.4151 - 0.3468i & 0.1863 - 0.1362i & 0.09796 - 0.06636i \\
 \hline
 \hline
\end{tabular}
\caption{QNMs accurate to 4 decimal places for $M=1$ scaled MOG gravitational perturbations $V_{i=2}^{(-)}$ for $\ell=2$, $\ell=3$ and $\ell=4$ 
modes.}
\label{Table2}
\end{table}

\pagebreak

In Table \ref{Table1} and Table \ref{Table2}, the QNMs were displayed for a scaled mass $M=1$ in order to illustrate the difference in magnitude with the result of general relativity. However, if we impose the same scaling condition $G_N M=1$ from GR to MOG we obtain \linebreak $GM=G_N(1+\alpha)M=1$ thus yielding $M=\frac{1}{1+\alpha}$. Consequently, we can observe that the QNMs will be increased by a factor of $(1+\alpha)$ and these will correspond to lower mass black holes than those predicted by GR. Table \ref{Table3} and Table \ref{Table4} show these results.

\begin{table}[h!]
\begin{tabular}{ |c|c|c|c|c|c| } 
 \hline
 \hline
 $\ell$ & $n$ & $\alpha = 0$ & $\alpha = 1$ & $\alpha = 4$ & $\alpha = 9$ \\ 
 \hline
 \multirow{4}{1em}{1} & 0 & 0.2483 - 0.09249i & 0.2896 - 0.09611i & 0.3171 - 0.09403i & 0.3268 - 0.09084i \\
   & 1 & 0.2145 - 0.2937i & 0.2616 - 0.3012i & 0.2941 - 0.2914i & 0.3038 - 0.2796i \\
   & 2 & 0.1748 - 0.5252i & 0.2271 - 0.5309i & 0.2629 - 0.5072i & 0.2675 - 0.4833i \\
   & 3 & 0.1462 - 0.7719i & 0.2179 - 0.7733i & 0.2518 - 0.7470i & 0.2434 - 0.7008i \\
 \hline
 \multirow{4}{1em}{2} & 0 & 0.4576 - 0.09500i & 0.5302 - 0.09833i & 0.5821 - 0.09650i & 0.6000 - 0.09351i \\
   & 1 & 0.4365 - 0.2907i & 0.5131 - 0.2995i & 0.5680 - 0.2927i & 0.5861 - 0.2830i \\
   & 2 & 0.4012 - 0.5016i & 0.4840 - 0.5126i & 0.5435 - 0.4974i & 0.5607 - 0.4791i \\
   & 3 & 0.3626 - 0.7302i & 0.4514 - 0.7397i & 0.5139 - 0.7125i & 0.5272 - 0.6840i \\
 \hline
 \multirow{4}{1em}{3} & 0 & 0.6569 - 0.09562i & 0.7542 - 0.09867i & 0.8239 - 0.09681i & 0.8493 - 0.09395i \\
   & 1 & 0.6417 - 0.2897i & 0.7418 - 0.2983i & 0.8136 - 0.2921i & 0.8391 - 0.2831i \\
   & 2 & 0.6138 - 0.4921i & 0.7189 - 0.5045i & 0.7944 - 0.4921i & 0.8198 - 0.4760i \\
   & 3 & 0.5779 - 0.7063i & 0.6891 - 0.7200i & 0.7687 - 0.6988i & 0.7927 - 0.6743i \\
 \hline
 \hline
\end{tabular}
\caption{QNMs accurate to 4 decimal places for $M=1/(1+\alpha)$ scaled MOG electromagnetic perturbations $V_{i=1}^{(-)}$ for $\ell=1$, $\ell=2$ and $\ell=3$ modes.}
\label{Table3}
\end{table}
\begin{table}[h!]
\begin{tabular}{ |c|c|c|c|c|c| } 
 \hline
 \hline
 $\ell$ & $n$ & $\alpha = 0$ & $\alpha = 1$ & $\alpha = 4$ & $\alpha = 9$ \\ 
 \hline
 \multirow{4}{1em}{2} & 0 & 0.3737 - 0.0890i & 0.4441 - 0.09300i & 0.5105 - 0.09333i & 0.5431 - 0.09171i \\
   & 1 & 0.3467 - 0.2739i & 0.4229 - 0.2847i & 0.4936 - 0.2839i & 0.5270 - 0.2781i \\
   & 2 & 0.3011 - 0.4783i & 0.3874 - 0.4914i & 0.4642 - 0.4848i & 0.4974 - 0.4725i \\
   & 3 & 0.2515 - 0.7051i & 0.3484 - 0.7158i & 0.4291 - 0.6985i & 0.4584 - 0.6776i \\
 \hline
 \multirow{4}{1em}{3} & 0 & 0.5994 - 0.0927i & 0.6706 - 0.09516i & 0.7479 - 0.09456i & 0.7875 - 0.09267i \\
   & 1 & 0.5826 - 0.2813i & 0.6563 - 0.2882i & 0.7360 - 0.2856i & 0.7761 - 0.2795i \\
   & 2 & 0.5517 - 0.4791i & 0.6298 - 0.4888i & 0.7138 - 0.4821i & 0.7543 - 0.4706i \\
   & 3 & 0.5120 - 0.6903i & 0.5957 - 0.7006i & 0.6842 - 0.6865i & 0.7238 - 0.6680i \\
 \hline
 \multirow{4}{1em}{4} & 0 & 0.8092 - 0.0942i & 0.8904 - 0.09608i & 0.9826 - 0.09513i & 1.030 - 0.09311i \\
   & 1 & 0.7966 - 0.2843i & 0.8795 - 0.2898i & 0.9735 - 0.2865i & 1.021 - 0.2802i \\
   & 2 & 0.7727 - 0.4799i & 0.8588 - 0.4881i & 0.9560 - 0.4812i & 1.004 - 0.4699i \\
   & 3 & 0.7398 - 0.6839i & 0.8301 - 0.6936i & 0.9315 - 0.6810i & 0.9796 - 0.6636i \\
 \hline
 \hline
\end{tabular}
\caption{QNMs accurate to 4 decimal places for $M=1/(1+\alpha
)$ scaled MOG gravitational perturbations $V_{i=2}^{(-)}$ for $\ell=2$, $\ell=3$ and $\ell=4$ 
modes.}
\label{Table4}
\end{table}
\pagebreak

By inspection of Table \ref{Table3} and Table \ref{Table4}, we can see that the damping given by Im$(\omega)$ of the QNMs for $\alpha=1$, $\alpha=4$ and $\alpha=9$ (corresponding to black holes that are $1/2$, $1/5$ and $1/10$ as massive, respectively, compared to its $\alpha=0$ counterpart) matches almost identically the damping predicted by GR. On the other hand, for increasing model parameter $\alpha$, the actual frequency of the QNM given by Re$(\omega)$ deviates and is significantly greater than that of GR.  This provides another key experimental signature of MOG in the ringdown phase.

\section{Conclusions}

We have investigated the quasinormal modes for gravitational and electromagnetic perturbations in the ringdown phase of the merging of two MOG black holes based on the STVG (MOG) theory. The black holes are MOG generalizations of the static, spherically symmetric Schwarzschild black hole in GR. For the parameter $\alpha > 0$ according to $G=G_N(1+\alpha)$, there is a significant reduction in the frequencies for MOG gravitational perturbations for $\ell=2, \ell=3$ and $\ell=4$ modes, and for $\ell=1, \ell=2, \ell=3$ MOG electromagnetic perturbations. This will allow for a possible distinguishing signal between MOG and GR for sufficiently sensitive frequency determinations. The detection of $\sim 50$ gravitational wave events by the aLIGO/Virgo observatories can be expected to produce accurate enough frequency results during the waveform ringdown phase. 

Attempts will be made in future work to calculate the MOG QNMs using the Kerr-MOG metric solution including the spin parameter $a$. It is anticipated that the spins of the merging black holes will play an important role when interpreting the aLIGO/Virgo ringdown data.

\section*{Acknowledgments}

Luciano Manfredi and Jonas Mureika thank the Perimeter Institute for Theoretical Physics for its generous hospitality during the initial stages of this research. They were respectively funded by a Summer Undergraduate Research Fellowship and Continuing Faculty Grant from the Frank R. Seaver College of Science and Engineering of Loyola Marymount University.  John Moffat thanks Luis Lehner and Martin Green for helpful discussions. This research was supported in part by Perimeter Institute for Theoretical Physics. Research at Perimeter Institute is supported by the Government of Canada through the Department of Innovation, Science and Economic Development Canada and by the Province of Ontario through the Ministry of Research, Innovation and Science.

\end{document}